\documentclass{aa}  
\usepackage{graphicx}
\usepackage{txfonts}
\bibpunct{(}{)}{;}{a}{}{,}

\usepackage{hyperref}
\hypersetup{
    colorlinks=true,
    citecolor=blue,
    linkcolor=blue,
    filecolor=magenta,
    urlcolor=cyan,
    pdftitle={main},
    pdfpagemode=FullScreen,
    }

\begin{document}

   \title{Which stars can form planets: Planetesimal formation at low metallicities}

   \author{Geoffrey Andama\inst{1}
          \and
          Jingyi Mah\inst{2}
          \and
          Bertram Bitsch\inst{3,2}
          }

   \institute{Department of Physics, Muni University, P.O. Box 725, Arua, Uganda\\
             \email{g.andama@muni.ac.ug}
             \and
             Max-Planck-Institut f\"{u}r Astronomie, K\"{o}nigstuhl 17, 69117 Heidelberg, Germany
             \and
             Department of Physics, University College Cork, Cork, Ireland
             }

   \date{Received 11 December 2023; accepted 26 January 2024}

  \abstract
   {The exoplanet diversity has been linked to the disc environment in which they form, where the host star metallicity and the formation pathways play a crucial role. In the context of the core accretion paradigm, the initial stages of planet formation require the growth of dust material from micrometre size to planetesimal size bodies before core accretion can kick in. Although numerous studies have been conducted on planetesimal formation, it is still poorly understood how this process takes place in low metallicity stellar environments. In this work, we explore how planetesimals are formed in stellar environments primarily with low metallicity. We performed global 1D viscous disc evolution simulations including growth of dust particles,  evaporation and condensation of chemical species at ice lines. We followed the formation of planetesimals during  disc evolution and tested different metallicities, disc sizes and turbulent viscosity strengths.  We find that at solar and sub-solar metallicities, there is a significant enhancement in the midplane dust-to-gas mass ratios at the ice lines but this leads to planetesimal formation only at the water ice line. In our simulations, [Fe/H] $= -0.6$ is the lowest limit of metallicity for planetesimal formation where a few Earth masses of planetesimals could form. Our results further show that for such extreme disc environments, large discs are more conducive than small discs for forming large amounts of planetesimals at a fixed metallicity, because the pebble flux can be maintained for a longer time resulting in a longer and more efficient planetesimal formation phase. At lower metallicities, planetesimal formation  is less supported in quiescent discs compared to turbulent discs which produce relatively larger amounts of planetesimals, because the pebble flux can be maintained for a longer time. The amount of planetesimals formed at sub-solar metallicities in our simulations places a limit on core sizes that could possibly only result in the formation of super-Earths. 
   }


   \maketitle

\section{Introduction}
It is indisputable that the population of over 5500 confirmed exoplanets is diverse in terms of their physical and chemical properties. Planets on close-in orbits have radii ranging from that similar to the Earth to that similar to Jupiter, with a noticeable gap between the super-Earth and sub-Neptune populations \citep[e.g.][]{Fulton2017}. In addition, for the gas giant population, recent observations have been making breakthroughs in characterising their atmospheric composition \citep[e.g.][]{Line2021,August2023,Bean2023,Pelletier2023}. In order to understand the origin of the diversity, it is crucial to understand the environment in which the planets form and the building blocks that are available when the planets were growing. These factors are inherently associated to the host star and its accompanying protoplanetary disc.

Statistical analyses of carefully selected samples of planets and their host stars reveal that stellar metallicity (commonly parametrised by [Fe/H]), among other stellar properties, exert a considerable influence on the occurrence rate of planets \citep[e.g.][]{UdrySantos2007,Petigura2018,Chen2022}. Giant planets, in particular, tend to be found around stars of super-solar metallicity \citep[e.g.][]{Santos2004,FischerValenti2005,Johnson2010,Mortier2013,Petigura2018}. This trend is also predicted by theoretical models of planet formation \citep[e.g.][]{IdaLin2004,Mordasini2012,Bitsch2015b,Ndugu2018}. Because stellar metallicity can be taken as a proxy for disc solid inventory \citep{HuehnBitsch2023}, discs around stars with higher metallicity tend to have higher mass \citep{Andrews2013} and metal content. In the case of favourable disc properties such as large disc radius and high disc viscosity, solids in the disc can be sustained for long periods \citep{BitschMah2023} which can facilitate the formation of large planets \citep{LambrechtsJohansen2014,SavvidouBitsch2023}. 

The correlation with stellar metallicity, albeit weaker, is also present for sub-Neptunes and super-Earths (\citealp{WangFischer2015,Petigura2018}; see however \citealp{Sousa2008,Buchhave2012}). Interestingly, the ratio of super-Earth to sub-Neptune increases with decreasing stellar metallicity \citep{Chen2022}, suggesting a correlation between stellar metallicity and the maximum size of the planetary cores that can form in the disc. 

In general, the diversity of planets hosted by stars with super-solar metallicity is of a wider range. For example, stars with super-solar metallicity have a higher probability of hosting close-in planets \citep{Mulders2016,Dong2018}, high eccentricity planets \citep{Buchhave2018,Mills2019}, and multiple planets in the case of M dwarf host stars \citep{RodriguezMartinez2023}. This is perhaps unsurprising because the more solid material available in the disc the more possible formation pathways that a planetary core can follow open up, which will lead to a variety of outcomes \citep[e.g.][]{Lambrechts2019}.

Planet occurrence rates around stars on the opposite end of the metallicity spectrum – the low-metallicity or iron-poor regime - have been relatively underexplored \citep{Mortier2012,Boley2021}. In models of planet formation, micrometre size dust first grows to pebbles via coagulation and condensation \citep{Zsom2010,Ciesla2010,Birnstiel2012,Ros2013} before overcoming a major growth barrier to become planetesimals. The formation of planetesimals is thought to proceed via the streaming instability (SI) mechanism \citep[e.g.][]{YoudinGoodman2005}. Studies have shown that SI is effective in regions of the disc where the local dust-to-gas mass ratio and the pebble size are sufficiently large \citep[e.g.][]{Carrera2015,Yang2017,LiYoudin2021}. \citet{DrazkowskaAlibert2017} propose that the water ice line is a prime location where the first planetesimals form. This relies on the change in pebble size across the water ice line which alters the drift speed of the pebbles, creating a ‘traffic jam’ that eventually triggers the gravitational collapse of dust clumps to form planetesimals. This location could also harbour a pressure perturbation driven by the difference in opacities caused by the difference in grain sizes interior and exterior to the water ice line \citep{Muller2021}. The ring of planetesimals at the water ice line would then continue to grow and evolve by collisions or pebble accretion \citep{BatyginMorbidelli2023,Woo2023}. 

Within the framework of this model, there should then be a lower limit in stellar metallicity below which the pebble flux at the water ice line is too low for planetesimal formation to kick in. Observations suggest that this limit is at $-0.6 < {\rm [Fe/H]} < -0.5$ \citep{Mortier2012} with the latest analysis suggesting that the limit could be lower \citep[between $-0.7$ and $-0.6$;][]{Boley2021}. Recent discoveries of planets around low-metallicity stars seem to be consistent with the metallicity limit \citep[e.g.][]{Hellier2014,Polanski2021,Brinkman2023,Dai2023}. 

Here we explore how planet formation works in a low metallicity environment from a theoretical point of view. We do this by simulating the viscous evolution of the gas and the growth of dust particles in a protoplanetary disc with a 1D model that includes an additional effect of evaporation and condensation of chemical species at ice lines \citep{SchneiderBitsch2021}. The evaporation and condensation effects generate a localised pile up of pebbles which can be converted into planetesimals under specific conditions \citep{DrazkowskaAlibert2017}. We focus on understanding the minimum requirements that permit planetesimal formation in such extreme environments and how various disc properties influence the location where planetesimal formation takes place, the efficiency with which planetesimals are formed and the total mass of planetesimals produced. 

We structure our work as follows. In Section~\ref{sec:methods}, we describe the disc and planetesimal formation models and the setup of the simulations. We present our results in Section~\ref{sec:results}, and then discuss and summarize our findings in Section~\ref{sec:discussion}.

\section{Methods}\label{sec:methods}
\subsection{Disc evolution and dust growth}
Our work uses the chemcomp code presented in \cite{SchneiderBitsch2021}. This 1D code computes the gas surface density by solving the viscous evolution equation \citep{LyndenBellPringle1974} and uses the alpha-viscosity prescription \citep{ShakuraSunyaev1073}.  The disc model used in this work is one with a smooth surface density profile without any pressure perturbations. The disc’s kinematic viscosity is expressed as
\begin{equation}
    \nu = \alpha_{\rm t} c_{\rm s}^2 \Omega_{\rm K}^{-1},    
\end{equation}
where $\alpha_{\rm t}$ describes the strength of turbulence in the disc, $c_{\rm s}$ is the isothermal sound speed and $\Omega_{\rm K}$ is the Keplerian orbital speed.

The temperature of the disc is contributed by two physical processes: viscous accretion (which operates in the inner disc) and stellar irradiation (which operates at larger orbital distances). The heat from viscous accretion depends on the disc's turbulent viscosity $(\alpha_{\rm t})$ as well as the initial dust-to-gas ratio (DTG). The heat from stellar irradiation, on the other hand, depends only on the stellar luminosity, which is a fixed value in this work. Thus, a disc with low DTG will have a lower temperature in its inner region compared to a disc with high DTG. For simplicity, the disc's temperature profile in the code does not evolve in time.

Dust grains in the disc start out initially as sub-micrometre-sized grains which then grow via coagulation into larger size grains called pebbles. The size of the pebbles is limited by radial drift (grains above the drift limit drift quickly towards the central star and are lost) and fragmentation (grains that collide with a relative velocity greater than the fragmentation limit are broken up into smaller fragments) \citep{Birnstiel2012}. Following results from laboratory experiments \citep{GundlachBlum2015}, we assume a fragmentation velocity of $u_{\rm f} = 10~{\rm m\,s}^{-1}$ for icy grains beyond the water-ice line and $u_{\rm f} = 1~{\rm m\,s}^{-1}$ for silicate grains closer in. The transition of fragmentation velocities results in different drift speeds of the grains at the water-ice line, leading to a ‘traffic jam’ (a region of high dust-to-gas ratio) just inside the water-ice line which could promote the formation of planetesimals \citep{DrazkowskaAlibert2017}.

The size and properties of dust grains in protoplanetary discs is a topic of intense research. Recent results from laboratory experiments \citep{MusiolikWurm2019} and simulations \citep{Jiang2023} suggest that icy dust grains may actually be fragile and thus fragment at lower velocities. We show the outcome of the scenario when the dust grains have fragmentation velocity of 1~${\rm m\,s}^{-1}$ in Appendix~\ref{sec:appendix_grain_size}.

A chemical model is also introduced to compute the composition of the dust grains. We show in Appendix~\ref{sec:appendix_disc_composition} the elements and molecules we included in our simulations and their respective abundances in the disc. As the pebbles drift inwards to the central star, they cross the ice lines of the different molecules in the disc and release the vapour of the corresponding molecules to the gas \citep{SchneiderBitsch2021}. Some of the vapour can diffuse outwards and recondense to make new pebbles. This effect results in a locally enhanced pebble surface density (and dust-to-gas ratio) at the location of the ice lines.

\subsection{Planetesimal formation}
We incorporate the planetesimal formation prescription of \citet{DrazkowskaAlibert2017}, in which planetesimals form in regions of the disc where the pebble Stokes number and dust-to-gas ratio in the midplane exceed $10^{-2}$ and 1, respectively. Pebbles are then converted to planetesimals at a rate of
\begin{equation}
	\Dot{\Sigma}_{\rm pla} = \zeta \cdot \Sigma_{\rm d} \cdot \Omega_{\rm K},
\end{equation}
where $\zeta$ is the efficiency and $\Sigma_{\rm d}$ is the dust surface density. We follow \citet{DrazkowskaAlibert2017} and use $\zeta = 10^{-3}$.

We modified the condition for the Stokes number according to recent results on the streaming instability \citep{LiYoudin2021}. In our model, planetesimals form when the following conditions are fulfilled:
\begin{itemize}
	\item the pebbles have Stokes number ${\rm St} \geq 10^{-3}$,
	\item the midplane dust-to-gas ratio $\varepsilon= \rho_{\rm d}/\rho_{\rm g} \geq 1$.
\end{itemize}
The Stokes number of the pebbles is computed dynamically in the code using
\begin{equation}
	{\rm St} = \frac{\pi}{2}\frac{a\rho}{\Sigma_{\rm g}},
\end{equation}
where $a$ is the pebble radius, $\rho$ is the pebble density and $\Sigma_{\rm g}$ is the gas surface density. As the code computes the vertically-integrated dust-to-gas ratio, we recover the midplane value using 
\begin{align}
	\varepsilon &=\frac{\rho_{\rm d}}{\rho_{\rm g}} = \frac{\Sigma_{\rm d}}{\Sigma_{\rm g}} \cdot \frac{H_{\rm g}}{H_{\rm d}},\\
	\varepsilon  &= \frac{\Sigma_{\rm d}}{\Sigma_{\rm g}}\sqrt{\frac{\alpha_{\rm z}+{\rm St}}{\alpha_{\rm z}}},
\end{align}
where $H_{\rm d}$ is the dust scale height, $H_{\rm g}$ is the gas scale height and $\alpha_{\rm z}$ describes the strength of vertical mixing. We note that \citet{DrazkowskaAlibert2017} did not make the distinction between $\alpha_{\rm t}$ and $\alpha_{\rm z}$. In our simulations, we found that planetesimals form when $\alpha_{\rm z} = 10^{-4}$, in agreement with the results of \citet{DrazkowskaAlibert2017}. However, we also tested the influence of vertical mixing in Appendix \ref{AppendixVerticalSettling}.

\subsection{Simulation setup}
In the first set of simulations, we study the conditions for planetesimal formation. We use a fixed mass for the central star $M_* = 1~M_{\odot}$ and varied the disc mass, the disc characteristic radius, the strength of viscous turbulence, and the stellar metallicity. The initial dust-to-gas ratio in the disc depends on the stellar abundance of all the elements included in the model. We computed this quantity using
\begin{equation}
    \label{eq:DTG}
    {\rm DTG} = \sum {\rm X/H} \cdot \mu_{\rm X} \cdot 10^{\rm [X/H]},
\end{equation}
where X/H is the initial abundance of element X in the disc (see Appendix~\ref{sec:appendix_disc_composition}), $\mu_{\rm X}$ is the atomic mass of element X, and [X/H] is the stellar abundance of element X as a function of [Fe/H] \citep{BitschBattistini2020}. For reference, we obtain DTG $= 0.016$ for [Fe/H] $= 0$.

In the second set of simulations, we explore the lower limit of the initial dust-to-gas ratio when planetesimal formation ceases. We fix the disc mass and disc radius to $M_{\rm disc} = 0.1~M_{\odot}$ and $r_{\rm c} = 100~{\rm au}$, respectively and varied $\alpha_{\rm t}$ and DTG. All simulations in both setups were run for 3~Myr with a time step of 10 years. Tables~\ref{tab:initial_conditions_set1} and \ref{tab:initial_conditions_set2} summarise the parameter spaces we explored.

 In the numerical setup, the inner and outer boundaries of the radial grid are set to 0.1 au and 1000 au, respectively, which allows all materials to leave our computational domain.
 
\begin{table}
    \caption{Parameter space explored in the first set of simulations.}
    \label{tab:initial_conditions_set1}
    \centering
    \begin{tabular}{lcc}
    \hline\hline
        Parameter & Symbol & Values\\
        \hline
        Stellar mass & $M_*$ & $1~M_{\odot}$ \\
        Disc mass & $M_{\rm disc}$ & 0.01, 0.05, $0.1~M_{\odot}$ \\
        Disc radius & $r_{\rm c}$ & 60, 100, 200~au \\
        Turbulent viscosity & $\alpha_{\rm t}$ & $10^{-4}$, $10^{-3}$ \\
        Vertical settling & $\alpha_{\rm z}$ & $10^{-4}$ \\
        Stellar metallicity & {\rm [Fe/H]} & $-0.4$, $-0.2$, 0.0 \\
        Dust-to-gas ratio & DTG & 0.008, 0.011, 0.016 \\
        \hline
    \end{tabular}
\end{table}

\begin{table}
    \caption{Parameter space explored in the second set of simulations.}
    \label{tab:initial_conditions_set2}
    \centering
    \begin{tabular}{lcc}
    \hline\hline
        Parameter & Symbol & Values \\
        \hline
        Stellar mass & $M_*$ & $1~M_{\odot}$ \\
        Disc mass & $M_{\rm disc}$ & $0.1~M_{\odot}$\\
        Disc radius & $r_{\rm c}$ & 100~au\\
        Turbulent viscosity & $\alpha_{\rm t}$ & $10^{-4}$, $10^{-3}$\\
        Vertical settling & $\alpha_{\rm z}$ & $10^{-4}$ \\
        Stellar metallicity & {\rm [Fe/H]} & $-0.7$, $-0.6$, $-0.5$ \\
        Dust-to-gas ratio & DTG & 0.005, 0.006, 0.007 \\
        \hline
    \end{tabular}
\end{table}

\section{Results}\label{sec:results}
\subsection{Planetesimal formation at ice lines in quiescent discs}
\begin{figure*}[h]
   \includegraphics[width=\textwidth]{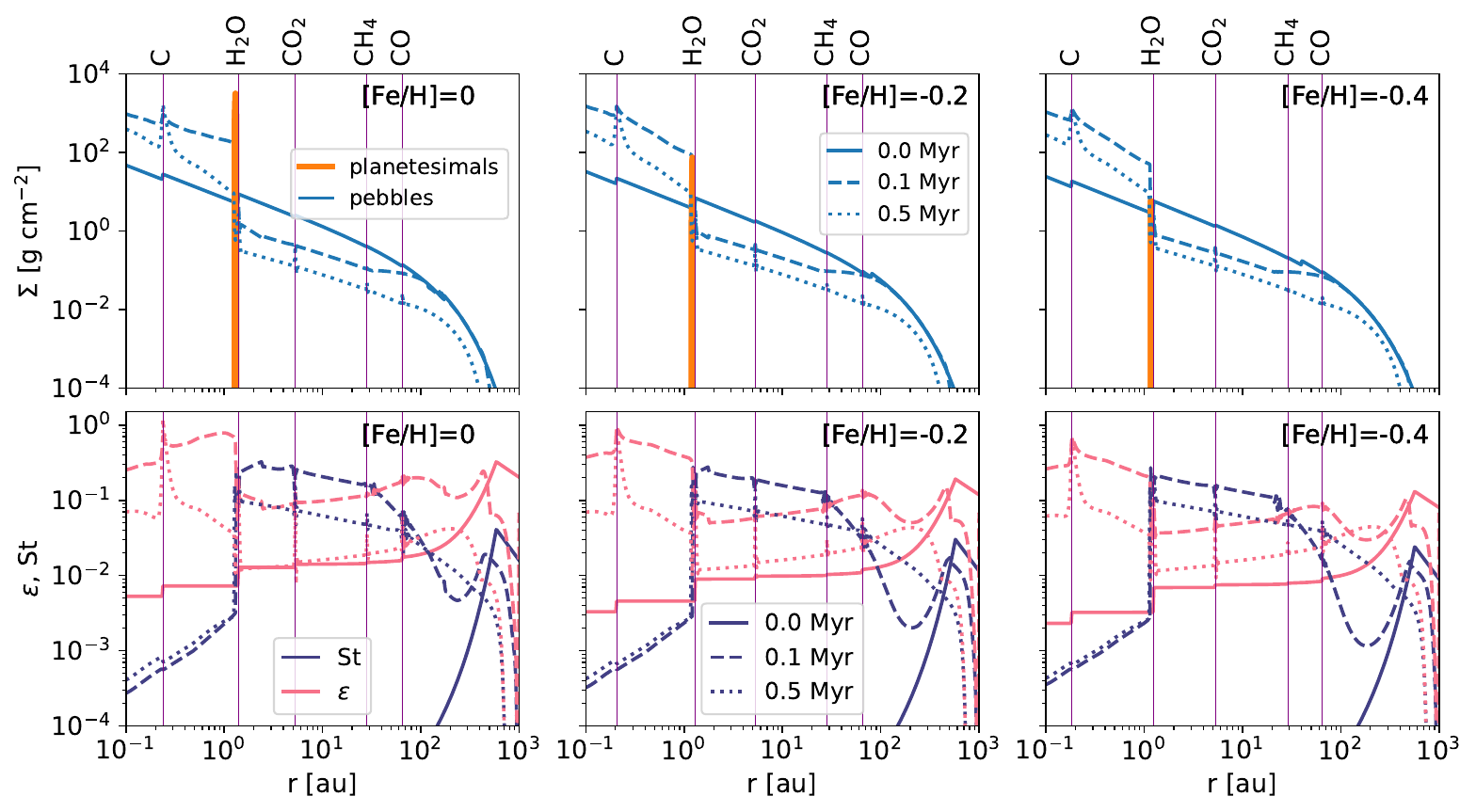}
   \caption{\textit{Top}: Evolution of pebble and planetesimal surface densities for a nominal $\alpha_{\rm t}=10^{-4}$ and disc size of $r_{\rm c}=100$ au and different values of disc metallicity. \textit{Bottom}: The evolution of the Stokes numbers and the corresponding midplane dust-to-gas ratios.}
   \label{fig:fig1}
\end{figure*}

\begin{figure*}[h]
   \includegraphics[width=\textwidth]{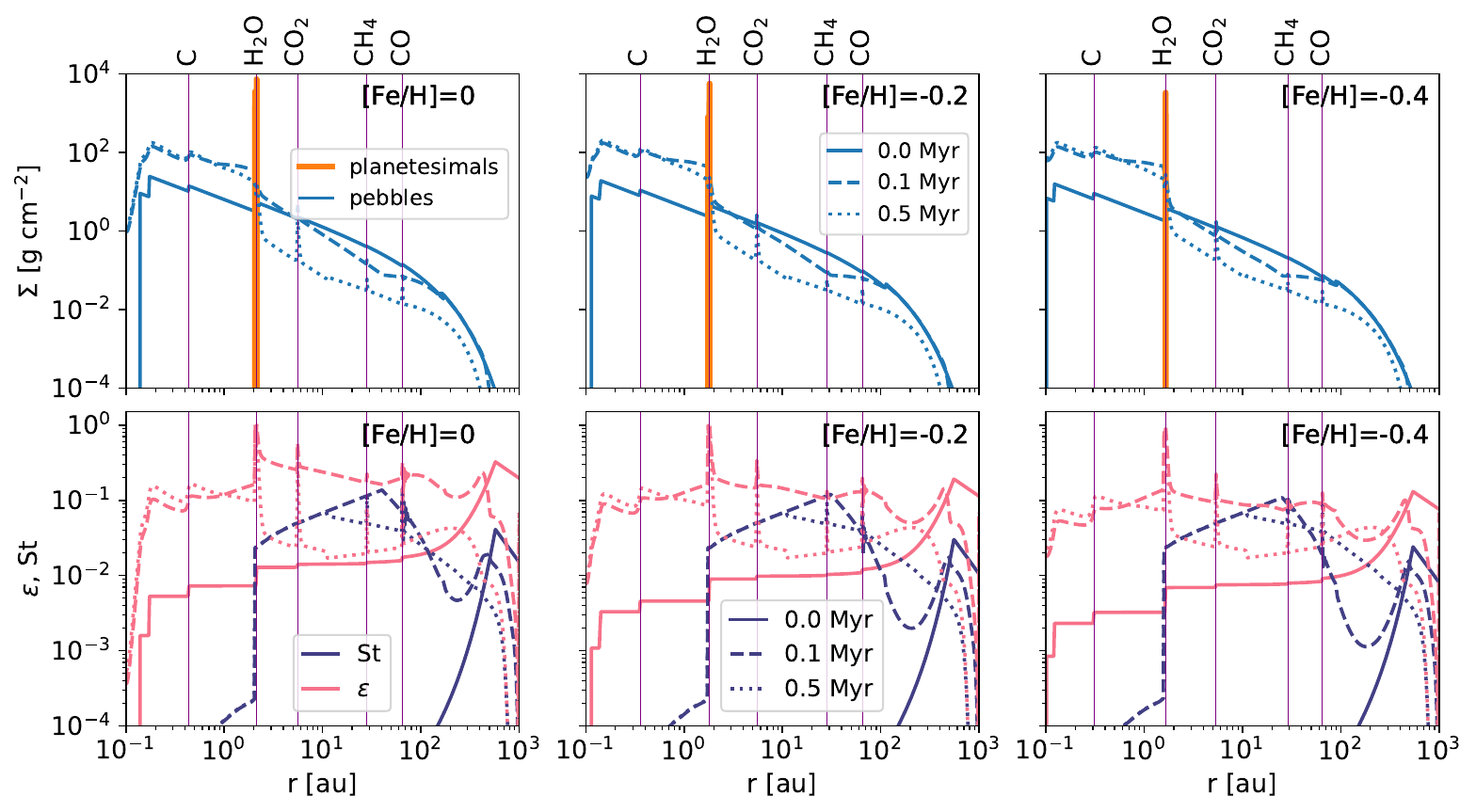}
   \caption{\textit{Top}: Evolution of pebble and planetesimal surface densities for a nominal $\alpha_{\rm t}=10^{-3}$ and disc size of $r_{\rm c}=100$ au and different values of disc metallicity. \textit{Bottom}: The evolution of the Stokes numbers and the corresponding midplane dust-to-gas ratios.}
   \label{fig:fig2}
\end{figure*}

In Figure \ref{fig:fig1}, we show our results for planetesimal formation in quiescent disc with turbulent viscosity of $\alpha_{\rm t}=10^{-4}$ at solar and sub-solar disc metallicities. In the top row, we show the evolution of the pebble and planetesimal surface densities at 0.1 Myr, 0.5 Myr and 1 Myr. Here, the lines for the planetesimal surface density at different times fall on top of each other because our planetesimal formation takes place within a narrow annulus within the ice line. However, the total mass evolution of the planestimals with time is provided in Section \ref{sec_mass_vs_time}.  In the bottom row, we show the connection between Stokes number and the midplane dust-to-gas ratio that set the conditions for planetesimal formation in our simulations.

As shown in the top panel of Figure \ref{fig:fig1}, planetesimal formation is prominent at the water ice line compared to the ice lines of the other chemical species. This can be explained by the interplay between the pebble Stokes number and the local dust-to-gas ratio  shown in the bottom panels of Figure \ref{fig:fig1}. Pebbles grow large in the less turbulent disc and possess drift limited sizes, which increase from outside to inside disc regions with a corresponding increase in the Stokes numbers. These large size pebbles drift rapidly to the water ice line region where they sublimate, releasing silicate grains. The released silicate grains undergo a transition in fragmentation velocities where they fragment at a lower velocity of 1 m/s in the disc regions interior to the water ice line, producing smaller size pebbles  with corresponding low Stokes number as shown in the bottom panel of Figure \ref{fig:fig1}. Consequently, these dust grains are more coupled to the gas and drift at reduced velocities compared to the large size pebbles in the disc regions exterior to the water ice line.  At the same time, the water vapour diffuses outward from the disc region interior to the ice line and re-condenses to make new pebbles at the ice line. The repetition of this process eventually leads to a traffic jam within the water ice line region \citep[see also][]{Pinilla2016, DrazkowskaAlibert2017}. The traffic jam effect causes a pile up of pebbles which leads to increased midplane dust-to-gas ratios as shown in the bottom panel of Figure \ref{fig:fig1}. In addition, due to the high Stokes numbers, pebbles decouple from the gas allowing them to settle, further boosting the midplane dust-to-gas ratio. However, in our simulations, the condition to make planetesimals is fulfilled within the water ice line region,  where no planetesimals form in the other parts as shown in the top panel  of Figure \ref{fig:fig1}. Here, planetesimals are formed at the water ice line because it is a sweet spot where we can have new pebbles formed by re-condensation of water vapor, but at the same time the Stokes numbers are still high enough to enable planetesimal formation. 

At the carbon sublimation line, pebbles also pile up because of the re-condensation of carbon grains as carbon vapour diffuses outward from the inside parts of the disc interior to the carbon sublimation line. Here, the dust pile up is stronger because the grains are small and do not drift so fast. This leads to high pebble surface densities and hence high dust-to-gas ratios at the carbon sublimation line as shown in the bottom panel of Figure \ref{fig:fig1}. However, because of the low Stokes numbers in this region, the pebbles are more coupled to the gas and are less settled which impedes planetesimal formation at the carbon sublimation line.

For the ice lines in the outer parts of the disc, the Stokes numbers are high but the dust-to-gas ratios are low, which makes it difficult for planetesimal formation to take place. This is because in these regions, the weak accumulation of pebbles at ice lines of the different chemical elements does not sufficiently enhance the dust-to-gas ratio to values suitable for planetesimal formation. 

We attribute the weak accumulation of pebbles at these ice lines to the fact that, in our simulations, there is no fragmentation velocity difference at the these ice lines that actually causes the traffic jam effect as at the water ice line. However, it needs to be tested if a fragmentation velocity transition could trigger planetesimal formation, especially at the CO$_2$ and CO ice lines, at solar and sub-solar metallicities. While the change in fragmentation velocity at the water ice line is motivated by experiments, it is difficult to perform the similar laboratory experiments for CO$_2$ and CO ice. However, numerical simulations seem to suggest that enhanced grain growth might not take place just outside the CO ice line \citep{Stammler2017}.

Although planetesimal formation appears to be difficult at the evaporation fronts other than the water ice line at solar metallicities, we show in Appendix \ref{AppendixSupersolar} the possibility that planetesimal formation  can take place at those ice lines in stellar environments with super solar metallicities. This is because at the super solar metallicities, significant quantities of frozen volatile species evaporate at the ice lines.  Hence, relatively larger amounts of vapour diffuses out compared to solar and sub-solar metallicities, which then re-condense to form new pebbles as explained before. In addition, at super solar metallicities, there is abundance of solid material that results in significant dust-to-gas ratio enhancements which promotes planetesimal formation \citep[e.g.,][]{LiYoudin2021}.

In the less turbulent discs, reducing disc metallicity to sub-solar values results in reduced planetesimal formation at the water ice line as shown in the middle and the right panels of Figure \ref{fig:fig1}. This is because, the lower the disc metallicity, the lower the amount of solid materials in the disc and the lower the dust-to-gas ratios. In addition, as already discussed above, dust grains drift faster in less turbulent discs (due to their larger size) which leads to faster depletion of solids and hence small amounts of solids accumulate and transform into planetesimals at the water ice line.

\subsection{Planetesimal formation at ice lines in turbulent discs}
 
Figure \ref{fig:fig2} shows how  planetesimal formation takes place in a turbulent disc environment (using $\alpha_{\rm t}=10^{-3}$). Again, here grain sizes play a key role for planetesimal formation. For example, in the bottom panel of Figure \ref{fig:fig2}, the grain sizes in the outer disc regions are in the drift limited regime. However, in the inner disc regions their sizes are limited by fragmentation levels. Hence, pebbles in the drift limited regime migrate  faster than the pebbles in the fragmentation limit (indicated by the radial drop in Stokes number interior to 10 - 30 au), which drift slower due to their smaller size. Here, the transition between drift and fragmentation limits shifts with time from the outer to the inner disc regions as shown in the bottom panel of Figure \ref{fig:fig2}. The slower drifting pebbles cause a traffic jam in the region interior to the point of transition between drift and fragmentation limits takes place. This results in accumulation of pebbles and hence more enhanced dust-to-gas ratio at the CO$_2$ line compared to the less turbulent disc as shown in the bottom panel of Figure \ref{fig:fig2}. However, both the Stokes number and the midplane dust-to-gas ratios at the  CO$_2$ line are below the thresholds that we set for planetesimal formation in our simulations.

In contrast to the weak turbulent viscosity, increasing the turbulence strength to $\alpha_{\rm t}=10^{-3}$ promotes planetesimal formation for the case of the sub-solar metallicities as shown in Figure \ref{fig:fig2}.
This is possibly because the high disc turbulence causes fragmentation of pebbles to relatively smaller sizes which drift at reduced speeds and hence linger in the disc for longer periods of time compared to quiescent disc \citep[see also][]{BitschMah2023}. On the other hand, the onset of the dust pile up at the ice line could also reduce the radial drift velocities of the dust grains \citep{Drazkowska2016}. The longer radial drift times give a window of opportunity for dust grains to concentrate at the water ice line, which can later form planetesimals once the planetesimal formation conditions are fulfilled. Here, we can maintain a high enough pebble flux for a longer time which in turn allows the formation of planetesimals for a longer time, resulting in a larger fraction of formed planetesimals.

Furthermore, for the case of high turbulence strength of $\alpha_{\rm t}=10^{-3}$, pebbles no longer accumulate at the carbon evaporation line, as shown in Figure \ref{fig:fig2}. This is because the high turbulence levels coupled with low fragmentation velocity in the inner disc regions converts  pebbles to smaller dust materials which have very small Stokes numbers as shown in the bottom panel of Figure \ref{fig:fig2}.   
These small size dust grain are more coupled to the gas and follow gas streamlines. These well-coupled dust grains drift more slowly but the high gas viscosity transports the carbon vapour away more efficiently onto the star. This impedes re-condensation of the carbon vapour to make new  pebbles at the carbon sublimation line, leading to relatively lower midplane dust-to-gas ratios compared to the less turbulent disc.

\subsection{The role of the volatiles}
The volatiles play an important role in regulating the dust-to-gas ratios and Stokes numbers as shown in the bottom panels of Figures \ref{fig:fig1} and \ref{fig:fig2}. For example, between the CO and water ice sublimation lines, the dust-to-gas ratios and the Stokes numbers gradually decrease and increase, respectively. At the sublimation fronts, there is a spike in both the dust-to-gas ratio and the Stokes number. 
In our simulations, particles grow to larger sizes as they migrate to the inner disc regions, where the Stokes numbers also increase.  Particles grow even larger at the sublimation lines as a result of the drift limited approximation, where the particle sizes increase with dust surface density \citep{Birnstiel2015}. The large size pebbles have correspondingly high Stokes numbers which lead to more efficient settling of pebbles, which in turn leads to an increase in the midplane dust-to-gas ratio. However, the increased Stokes number also allows pebbles to drift faster to the next ice line, which leads to a drop in the dust-to-gas ratios between the ice lines. Here, pebbles instead accumulate temporarily, especially at ice lines for the volatile pebble species.

The accumulation of pebbles is more pronounced within the water ice line because there is a change in fragmentation velocities of the grains interior and exterior to the ice line region. As mentioned before, the larger ice rich pebbles drift fast, increasing the water vapor in the inner disc interior to the water ice line.  In the interior disc regions, the released dry silicates remain small in size because they fragment at a low velocity of 1 m/s. The small dust grains have corresponding low  Stokes numbers interior to the water ice line.
The small dust grains then become more coupled to the gas and the outward diffusion carries them outward where they can grow back to larger sizes leading to accumulation of pebbles within the ice line region.  This back and forth grain transport process is enhanced by increasingly rapid inward drift of pebbles in-between the ice lines. The rapid transition of the drift velocity at the water ice line leads to a traffic jam, high accumulation of pebbles and increased dust-to-gas ratios which promote planetesimal formation \citep{DrazkowskaAlibert2017}.

A change in fragmentation velocity at different ice lines of the volatile pebble species could result in dust surface density enhancements \citep{Pinilla2017} leading to possibility of planetesimal formation at those ice lines. In addition, studies show that sintering of aggregates consisting of multiple species of volatile ices could result in dust pile up in multiple locations close to the ice lines \citep{Okuzumi2016}. Consequently, this could also support planetesimal formation at the different ice lines under favourable conditions. However, the effect of both sintering and changes in the fragmentation velocity at multiple ice line locations is beyond the scope of this paper and will be treated in a future work.

\subsection{Time evolution of planetesimal mass at water ice line}\label{sec_mass_vs_time}
To put planetesimal formation at the water ice line in perspective in terms of total mass of pebbles  converted into planetesimals, we explored four parameters that might constrain planetesimal formation in our disc model. We then discuss what time planetesimal formation kick-starts during disc evolution.
\subsubsection{Dependency on metallicity}
\begin{figure}
   \includegraphics[width=0.5\textwidth]{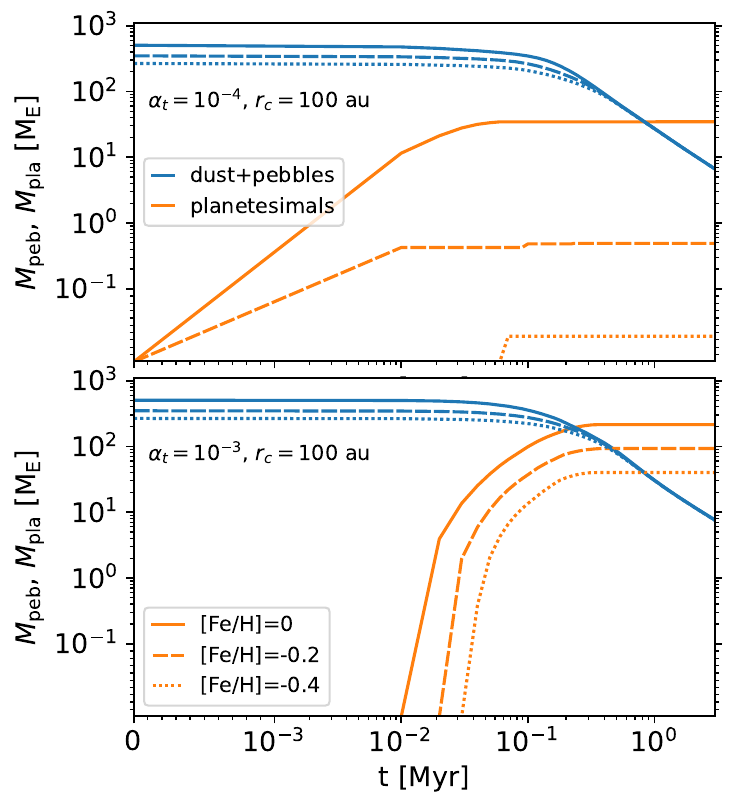}
   \caption{Metallicity dependency of planetesimal formation at the water ice line.}
   \label{fig:fig3}
\end{figure}
\begin{figure}
   \includegraphics[width=0.5\textwidth]{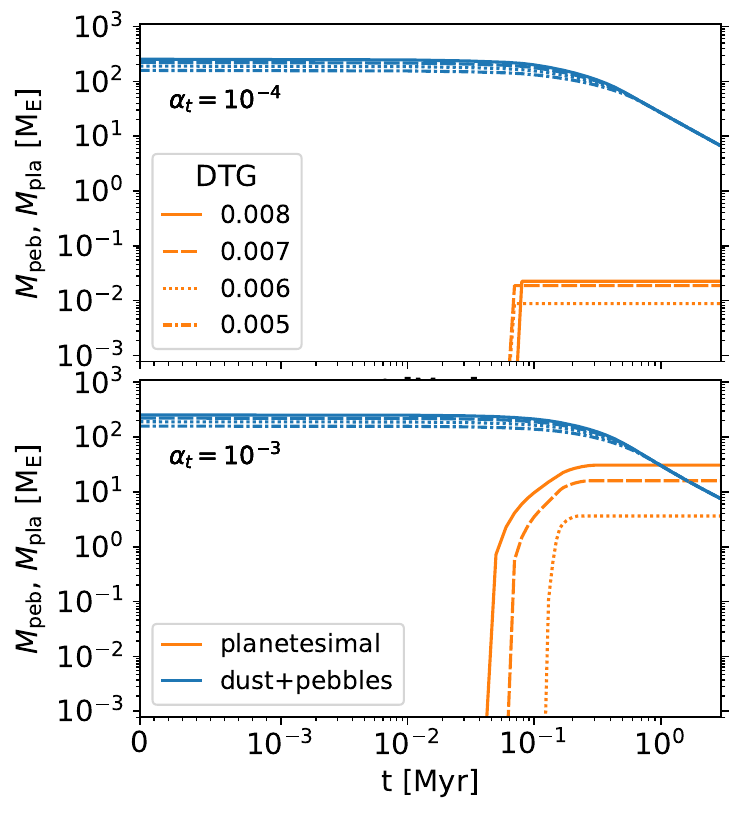}
   \caption{The lower limit of dust-to-gas ratio for planetesimal formation at the water ice line in a disc with $r_{\rm c}=100$ au.}
   \label{fig:fig7}
\end{figure}

We explored planetesimal formation for different values of disc metallicity ranging from solar to sub-solar values to infer the minimum requirement of disc metallicity for planetesimal formation. Our results are shown in Figure \ref{fig:fig3}, where the general trend is that the amount of planetesimals in Earth masses formed decreases as the disc metallicity decreases. This is expected because the disc metallicity, which is used as a proxy for calculating the dust-to-gas ratio, determines  the availability of solid materials in the disc. Which means, the higher the metallicity, the higher the amount of solids available for potential formation of planetesimals.  

The top panel of Figure \ref{fig:fig3}  shows how planetesimal formation takes place in a quiescent disc at different metallicity values.  Here, planetesimals form within 0.05 Myr of disc evolution, where no more conversion of pebbles takes place shown by the flat curves.  Here, a few Earth masses of up to 35 $M_{\rm{E}}$ of planetesimals form for the case of [Fe/H] $= 0$.  However, the amount of planetesimals formed drops to very low values when we reduce the metallicity to sub-solar values. For example, for the case of [Fe/H] $=-0.4$, very negligible amount of planetesimals is formed and only a fraction of Earth masses of planetesimal form for the case of [Fe/H] $=-0.2$. This suggests that a quiescent disc is not a favourable environment for formation of planetesimals at low disc metallicities. Probably, this is because, at low viscosity, the grains interior to the ice line are still large enough that they slightly decouple from the gas and drift inwards fast enough to prevent the traffic jam effect. In contrast, at higher viscosity, the traffic jam effect persists, because the pebbles are smaller and thus drift inward slower as explained earlier \citep[see also][]{BitschMah2023}.

In the turbulent disc with $\alpha_{\rm t}=10^{-3}$, as shown in the bottom panel of Figure \ref{fig:fig3}, considerable Earth masses of planetesimals are formed even at low disc metallicities, where at least 40 $M_{\rm{E}}$ planetesimals form for the case of [Fe/H] = $-0.4$. This is several orders of magnitude larger compared to a quiescent disc with a metallicity of [Fe/H] $= -0.4$.  This demonstrates that moderately turbulent disc environments are beneficial for planetesimal formation because of their grain retention capabilities as earlier explained. However, in highly turbulent discs, for example with $\alpha_{\rm t}=10^{-2}$, grains might become too small in the fragmentation limit and drift too slowly that planetesimal formation becomes difficult \citep[e.g.,][]{DrazkowskaAlibert2017}.  Nevertheless, for planetesimal formation to take place at the water ice line in stellar environments with sub-solar metallicities, the surrounding discs must be turbulent enough in order to produce relatively small size pebbles and scale down the loss of pebbles via radial drift on short dynamical timescales. However, at low viscosities, the presence of pressure bumps could also slow down grain migration \citep{Pinilla2012, Andama2022}, which could allow more planetesimals to form at the ice line \citep[e.g.,][]{Stammler2019}.

Our numerical code lacks scaling for metallicity values below [Fe/H] $=-0.4$ that self-consistently computes dust-to-gas ratios for dust evolution and the accurate chemical composition.  This is because the stellar abundances from GALAH surveys \citep{Buder2018} on which the dust-to-gas scaling in the code is based \citep{BitschBattistini2020, SchneiderBitsch2021} span [Fe/H] values from $-0.4$ to $+0.4$. However, we extrapolated the dust-to-gas ratios to the lowest value of [Fe/H] $=-0.7$.  Therefore, in order to infer the lowest limit placed on metallicity for planetesimal formation, we performed simulations directly with lower values of dust-to-gas ratio below the value for  [Fe/H] $=-0.4$, ranging from 0.005 to 0.008, corresponding to [Fe/H] $=-0.7$ and [Fe/H] $=-0.4$, respectively. The results are shown in Figure \ref{fig:fig7}, which suggest that a dust-to-gas ratio of about 0.006 (corresponding to [Fe/H] $=-0.6$) is the lower limit required for planetesimal formation to take place. In our simulations, for this lower limit of dust-to-gas ratio, only a Moon mass amount of planetesimals can form in a quiescent disc, but a few Earth masses of planetesimals can form for the case of a turbulent disc.  Here, as shown in Figure \ref{fig:fig7}, planetesimal formation starts relatively late because  the amount of disc material is small, which takes time to accumulate at the ice lines in contrast to discs with higher metallicities. Our results on the limits placed by the availability of dust materials for planetesimal formation to take place are in line with the results of the recent streaming instability simulations \citep{LiYoudin2021}.

\subsubsection{Dependency on disc mass}
\begin{figure}
   \includegraphics[width=0.5\textwidth]{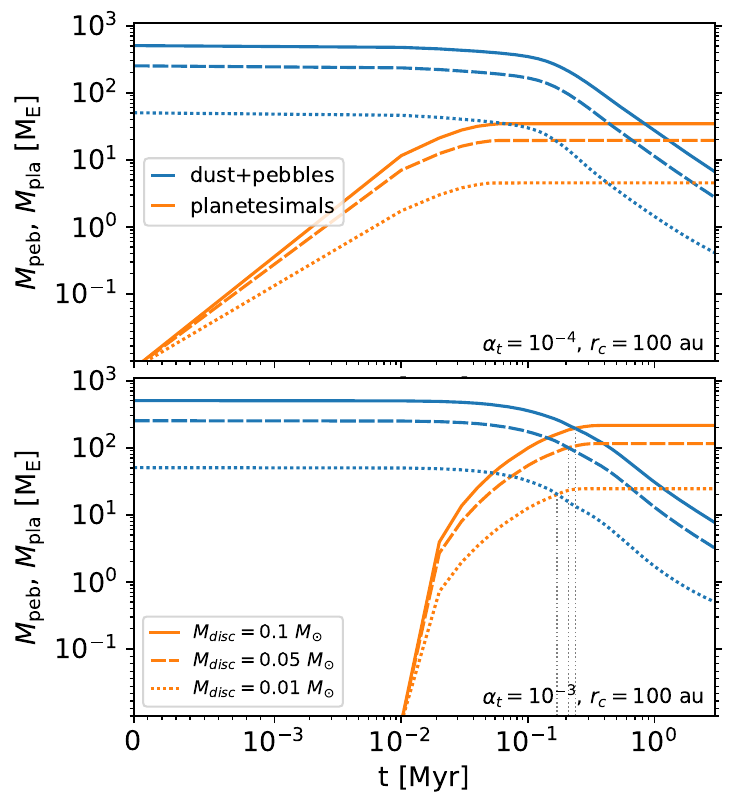}
   \caption{Disc mass dependency of planetesimal formation at the water ice line in a disc environment with [Fe/H] $= 0$. The vertical dotted lines in the bottom panel show the points when the curves start to level off, where planetesimals stop forming.}
   \label{fig:fig4}
\end{figure}

In Figure \ref{fig:fig4}, we show how planetesimal formation at water ice line scales with the initial disc mass and turbulent viscosity, where  the top and bottom panels show quiescent and turbulent discs respectively. To test the lower limit of disc mass that can support planetesimal formation at the water ice line, we performed simulations with lower disc masses $M_{\rm{disc}}= 0.05$ $M_{\odot}$ and $M_{\rm{disc}}=0.01~M_{\odot}$, with the corresponding initial dust masses of 250 $M_{\rm{E}}$ and 50 $M_{\rm{E}}$, respectively.

As shown in Figure \ref{fig:fig4}, relatively significant amounts of planetesimals are formed even in the least massive disc in our simulations, where we obtained at least 25 $M_{\rm{E}}$ in planetesimals for the case of $M_{\rm{disc}}=0.01 M_{\odot}$ and $\alpha_{\rm t}=10^{-3}$. Here, nearly half of the pebbles are lost to the central star via radial drift. This brings to question whether pebble accretion can support planet formation in stellar environments with such a low disc mass, which is in fact difficult as shown in \cite{SavvidouBitsch2023}. However, the above picture may not necessarily limit pebble accretion. For example, substantial amounts of planetesimals are formed before 0.1 Myr, where the formation process stops at around 0.2 Myr. These planetesimals could coalesce into a planetary embryo which could accrete any subsequent pebbles that accumulate at the ice line. Moreover, the planetary embryo at the water ice line could accrete the accumulated pebbles very efficiently before they are converted into planetesimals \citep{Morbidelli2020,Izidoro2021, Andama2022} or accrete other planetesimals within this region \citep{Chambers2023, BatyginMorbidelli2023}.

In summary, within the limits of disc masses explored in this studies, the existence of a water ice line within a disc with a transition in fragmentation velocity at the ice line is sufficient to promote planetesimal and hence planet formation. Which means even stellar environments with low disc masses are still conducive for planet formation, provided they are sufficiently turbulent with large radial extent to minimize rapid loss of pebbles.

\subsubsection{Dependency on disc size}
\begin{figure}
   \includegraphics[width=0.5\textwidth]{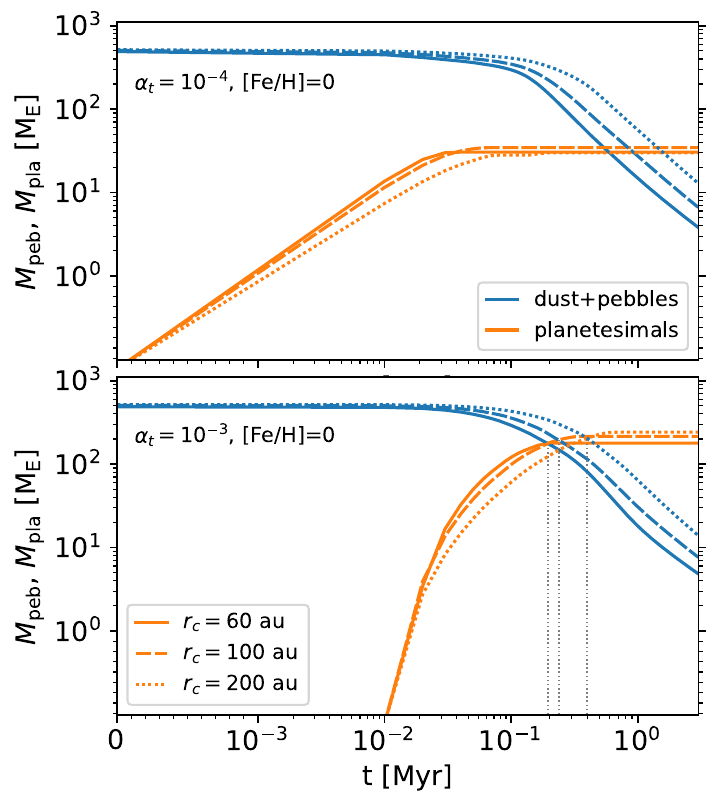}
   \caption{Disc size dependency of planetesimal formation at the water ice line. The vertical dotted lines have the same meaning as in Figure \ref{fig:fig4}.}
   \label{fig:fig5}
\end{figure}

We ran another set of simulations to test the dependency of planetesimal formation on the disc size, and our results are shown in Figure \ref{fig:fig5}. In the  quiescent discs as shown in the top panel of Figure \ref{fig:fig5}, the planetesimal formation times at the water ice line are insensitive to the disc size, where roughly the same amount of planetesimals take about the same time to form. On the other hand, in the bottom panel of Figure \ref{fig:fig5} for the turbulent disc, planetesimal formation takes longer time in a large disc compared to a small disc. 
These observed trends can best be explained by how the disc size affects pebble flux during the course of disc evolution. In large discs, we can maintain pebble flux for a longer time than in small discs, which is, however, smaller, explaining the initially slower planetesimal formation efficiency at the water ice line compared to smaller discs, where the pebble flux is initially higher. However, for a low viscosity of $\alpha_{\rm t}=10^{-4}$, we lose pebbles faster, almost on similar dynamical time scales, in both large and small discs with small difference in the amount of formed planetesimals.

In turbulent discs with $\alpha_{\rm t}=10^{-3}$, the completion times of planetesimal formation at the water ice line depend on the radial extent of the disc, which increases with increasing $r_{\rm c}$. For example, in our simulations, it takes about 0.2 Myr  and  0.4 Myr for planetesimal formation at the water ice line to complete in a small disc with $r_{\rm c}=60$ au and a large disc with $r_{\rm c}=200$ au, respectively. The reason for this is that, at high viscosity, pebbles take longer time to drift from the outer disc regions and accumulate at the water ice line. Consequently, in larger discs, pebbles drift over longer distances compared to the smaller size discs, where the drift times are shorter. In addition, it takes longer for the pebble flux to decay in larger discs, because the pebbles still need to form in the outer regions before they drift inwards \citep{LambrechtsJohansen2014, Bitsch2018, BitschMah2023}.

From the bottom panel of Figure \ref{fig:fig5}, about 250 $M_{\rm{E}}$ of planetesimals form at the water ice line in the large disc with $r_{\rm c} = 200$ au compared to the small size disc with $r_{\rm c} = 60$ au in which roughly 190 $M_{\rm{E}}$ of planetesimals form. This is because, as mentioned above, the large disc retains pebbles for longer periods as more pebbles are produced from the disc's outskirts with the outward shift in the pebble production line \citep{LambrechtsJohansen2014}. In contrast, in small discs, pebbles drain faster and hence planetesimal formation stops early on, resulting in small amounts of planetesimals. In our simulations, planetesimal formation stops before 1 Myr, inconsistent with the Solar system, where planetesimal formation is expected to take place at all stages, as inferred from the estimates of chondrite ages \citep{Amelin2002,Kleine2009,Connelly2012,Kruijer2014,Kruijer2017,Onyett2023}. However, this can be attributed to the disc structure used in our simulations that do not feature disc substructures due perturbations in the disc structure. As mentioned earlier, perturbations in the gas density could accumulate pebbles and aid planetesimal formation at later stages of disc evolution \citep[e.g.,][]{Izidoro2021}.

\subsubsection{Dependency on disc turbulence}
\begin{figure}
   \includegraphics[width=0.45\textwidth]{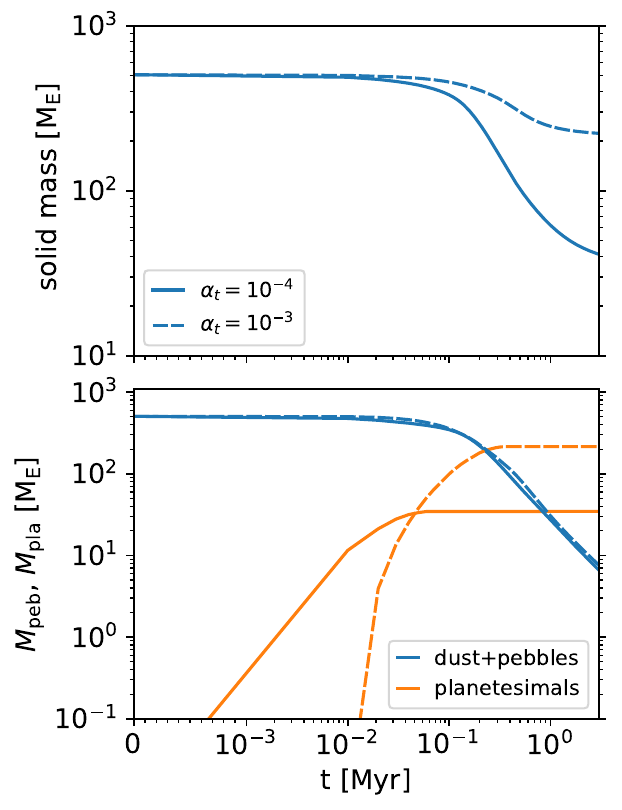}
   \caption{Disc turbulence dependency of planetesimal formation at the water ice line in a disc with $r_{\rm c}=100$ au and [Fe/H] $= 0$. {\textit{Top}}: The time evolution of the total solid mass without planetesimal formation taking place in the disc. {\textit{Bottom}}: The time evolution of dust and pebble mass including planetesimal formation in the disc. }
   \label{fig:fig6}
\end{figure}
The main role played by the turbulent viscosity has been discussed in the previous sections. Here, we further show how it impacts planetesimal formation in terms of the total dust mass lost via radial drift of pebbles for nominal disc size of $r_{\rm c}=100$ au and disc metallicity of [Fe/H] $= 0$. As shown in the bottom panel of Figure \ref{fig:fig6}, about 35 $M_{\rm{E}}$ of the intial dust mass of about 400 $M_{\rm{E}}$ is converted into planetesimals at the water ice line in a quiescent disc with $\alpha_{\rm t}=10^{-4}$. In contrast, for the case of a turbulent disc with $\alpha_{\rm t}=10^{-3}$, about 200 $M_{\rm{E}}$ of planetesimals are formed, representing 50\% conversion rate of the initial total solid mass. As shown in the top panel of Figure \ref{fig:fig6}, after 3 Myr of disc evolution about 200 $M_{\rm{E}}$ of dust is still present in the disc with $\alpha_{\rm t}=10^{-3}$. However, for the disc with $\alpha_{\rm t}=10^{-4}$, about 30 $M_{\rm{E}}$ of dust material remains. As discussed before, dust grains are depleted more rapidly in quiescent discs compared to the turbulent discs, which have better dust retention, which explains the disc mass evolution trends in Figure \ref{fig:fig6}.

In addition, disc turbulence may impact planetesimal formation at the ice lines by influencing the diffusive processes within the ice line regions. For example, high turbulence levels could transport material outward to the ice line regions faster (compared to low turbulence levels) where dust grains can grow to centimetre size pebbles \citep{Ros2013, Ros2019}. However, in our simulations, we did not test to what extent this process would affect planetesimal formation at the ice lines.

\subsection{Comparison with other studies}

In our simulation, we record accumulation of pebbles within a very narrow annulus around the ice line in contrast to \citet{DrazkowskaAlibert2017}. This is because we did not include backreaction of the solids on the gas that would cause traffic jam effect in broader disc regions. In addition, in the \citet{DrazkowskaAlibert2017} model, both the changes in fragmentation velocity at the ice line and the dust backreaction on the gas work in tandem, further amplifying the traffic jam effect. However, in our simulations, the traffic jam is solely driven by the changes in fragmentation levels of the pebbles at the ice line. Without backreaction, we were able to  form planetesimals in a broader disc region after increasing the dust-to-gas ratio to at least 3\% (see Appendix \ref{AppendixSupersolar}).

\citet{DrazkowskaAlibert2017} included a temperature evolution and perturbations around the ice line, using the \cite{Bitsch2015a} disc model, which further makes planetesimal formation more efficient than in our model. Firstly, the perturbations around the ice line region delay pebble flux and hence more planetesimals are formed. Secondly, the region where planetesimals are formed is larger due the disc temperature evolution which shifts the ice line region, which we did not include in our simulations.

\citet{DrazkowskaAlibert2017} focused on planetesimal formation at the water ice line in contrast to our model which features several evaporation fronts. Hence, our model sheds more light on dust pile up and hence the possibility of planetesimal formation at other ice lines other than the water ice line. 

Another key difference between our results and that of \citet{DrazkowskaAlibert2017} is the metallicity effect, where the authors reported a difficulty in forming planetesimals at water ice line at solar metallicities unless the disc in non-irradiated. In the other disc models the authors tested, rather high dust-to-gas ratio are needed to form planetesimals at the water ice line. In contrast, our simulations reveal the minimum dust-to-gas ratios far below the solar values that support planetesimal formation at the water ice line.
This could be because we included particles with Stokes number larger than 0.001, while \citep{DrazkowskaAlibert2017} only included particles with St > 0.01, which could also enhance our planetesimal formation efficiency at low metallicities.

Despite the inherent differences discussed above, both our model and the \citet{DrazkowskaAlibert2017} model have highlighted the sensitivity of planetesimal formation, especially at the ice line, to the disc parameters.

Other previous studies with results similar to ours include \cite{Schoonenberg2018} 
 and \cite{Kalyaan2023}, where planetesimal formation takes place at water ice line as a result of traffic jam effect experienced by pebbles. However, in \cite{Schoonenberg2018}, planetesimals could form in broad areas interior to the water ice line in contrast to our results where planetesimals formed only at the water ice line. This could be due to the fact that in \cite{Schoonenberg2018}, the authors used a fragmentation velocity of 3 m/s for the dry silicates which results in relatively larger dust grains compared to the small grains interior to the water ice line in our simulations. The relatively large dust grains have higher Stokes numbers, which means they can easily reach the threshold for planetesimal formation. The major difference between our work and that of \cite{Kalyaan2023} is that, in their study, planetesimals were able to form at multiple locations in the outer disc regions primarily because the author included pressure bumps in their model, which allows an accumulation of particles and thus enhanced planetesimal formation \citep[e.g.,][]{Johansen2007}. 

\section{Discussion and summary}\label{sec:discussion}

In this work, we explored a range of disc parameters to study how planetesimal formation might take place at ice lines driven by drifting and evaporating pebbles by performing full viscous disc evolution. Our key results have unveiled the minimum requirements of disc properties suitable for  the formation of planetesimals, especially at solar and sub-solar metallicities.

In our simulations, we could not form planetesimals interior to the water ice line despite the substantial accumulation of pebbles at the evaporation lines in the inner disc regions. Nevertheless, although the accumulated pebbles might not form planetesimals, the enhanced dust-to-gas ratios at the evaporation fronts could be beneficial for pebble accretion when a growing core accretes at those locations. This could be a possible formation pathway for hot and dense Jupiters around M dwarfs and also be useful for understanding the origin of heavy element content of certain exoplanets including Jupiter.

In addition to the water ice line, planetesimal formation can take place at the ice lines of the volatile species in the outer disc regions.  However, it requires super solar metallicites for planetesimals to form at those ice lines (see Appendix \ref{AppendixSupersolar}). Hence, in addition to supporting planet formation,  the planetesimals formed at the ice lines  in the outer regions could be source of planetesimal disc such as those in the Kuiper belt in the Solar system \citep[e.g.,][]{Eistrup2019}. Additionally, Jupiter's core could have formed as a result of the presence of the water ice line \citep{Walsh2011,Savvidou2021}, although other formation pathways could also play a major role.

As shown in this study, the  formation of planetesimals at the ice line is a spontaneous process once the Stokes number and the dust-to-gas ratio thresholds necessary for planetesimal formation are reached. This  process can take place in the very early stages of disc evolution, which sets the stage for early formation of planetary bodies \citep[e.g.][]{Manara2018, SavvidouBitsch2023}. In quiescent discs, planetesimal formation at the ice line is ephemeral because dust grains grow large and are lost more quickly via radial drift. In contrast, turbulent discs have better dust retention capability and as such, the planetesimal formation process continues until the pebble flux drops below the required thresholds for enhancing dust-to-gas ratio at the water ice line. 

Furthermore, we have shown that planetesimal formation is a ubiquitous process that could take place in diverse disc environments  purely driven by the presence of the ice lines, including those stellar environments with sub-solar metallacities and an initial dust composition of at least 0.6\% corresponding to [Fe/H] $= -0.6$. However, only a few Earth masses of planetesimals form in such metal-poor stellar environments which could also explain why the occurrence rate of super-Earths drops around [Fe/H] $= -0.5$ \citep{BashiZucker2022}.

Although originally the drift barrier is thought to lead to rapid loss of solid material to the central star, this formidable process appears to be a holy grail for planetesimal formation at the water ice line. Hence the centimetre size pebbles might not be completely lost via rapid radial drift as previously thought since a substantial amount is instead transformed into planetesimals at the water ice line. Notably, in our simulations, this process occurs in the early stages of disc evolution which is beneficial for planet formation at the water ice line because the formed planetesimals could subsequently accrete the incoming pebbles. However, because of the low fragmentation levels, it is very hard to make planetesimals in the inner disc (interior to the ice line). This is a problem, especially for terrestrial planet formation and rocky super-Earth formation unless their formation starts at or beyond the water ice line where they could then migrate inward. However, this problem could be circumvented by a pressure perturbation which can trap pebbles, leading to planetesimal formation, and later pebble accretion in the interior disc regions.

The disc regions interior to the water ice line exhibit high dust to gas ratios but very low Stokes numbers, which is again a problem for pebble accretion in this region. First of all, the very low Stokes numbers make pebble accretion inefficient, which could however be accreted along with the gas instead once the planets have reached the gas accretion stage \citep{Morbidelli2023, BitschMah2023}. Secondly, the dust grains in these region might be just the left over pebbles that could not be converted into planetesimals at the ice line. This might be possible especially if the ice line acts as an efficient site for planetesimals formation where most of the pebbles are converted into planetesimals. 

Our simulations lacked detailed grain dynamics that includes backreaction of the dust material on the gas and different fragmentation levels which might not necessarily be as low as 1 m/s as here assumed interior to the water ice line. Thus, this could significantly change the picture of planetesimal formation in the interior disc regions. Nevertheless, our results have pointed out a central role played by the evaporation fronts in planetesimal formation and consequences for the formation of terrestrial and rocky planets.
Although we focused mainly on planetesimal formation at the different ice lines,  planetesimal formation can take place elsewhere in the disc, for example, in zonal flows \citep{Johansen2011, Dittrich2013}, in vortices \citep{Raettig2015, Surville2016} and at the edges of the dead zone \citep{Lyra2009}. However, in the likelihood that the conditions for these mechanisms do not exist, the ice lines could be an ideal location for planetesimal formation and hence pebble accretion.

\begin{acknowledgements}
We thank the anonymous referee for the valuable comments, which helped us improve the quality of our paper.
G.A. acknowledges the support from the Max Planck Society's ARTEMIS -- "African Research Talents Experiencing Mentorship In Science" program.
J.M. and B.B. acknowledge the support of the DFG priority program SPP 1992 ``Exploring the Diversity of Extrasolar Planets'' (BI 1880/3-1). B.B. thanks the European Research Council (ERC Starting Grant 757448-PAMDORA) for their financial support.
\end{acknowledgements}

    \bibliographystyle{aa}
    \bibliography{Andama_Mah_Bitsch_2023}

\begin{appendix} 
\section{Influence of grain size on planetesimal formation}
\label{sec:appendix_grain_size}

\begin{figure*}
   \includegraphics[width=\textwidth]{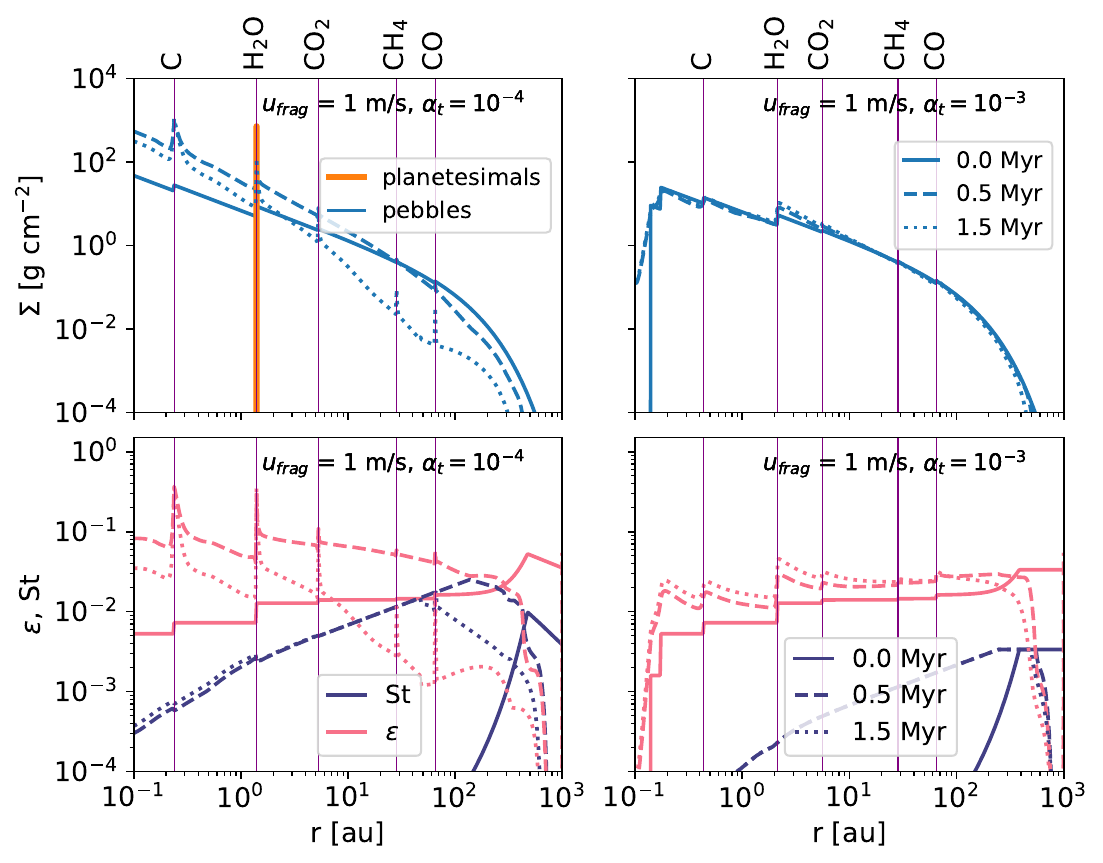}
   \caption{Planetesimal formation in a disc with a fixed grain fragmentation velocity of $u_{\rm{frag}}=1~\rm{m/s}$, solar metallicity [Fe/H] $= 0$, $M_{\rm disc} = 0.1~M_{\odot}$ and a disc size $r_{\rm c}=100$ au. {\textit{Left}}: A quiescent disc with $\alpha = 10^{-4}$. {\textit{Right}}: A turbulent disc with $\alpha = 10^{-3}$.}
   \label{fig:figA0}
\end{figure*}

 In view of recent results which hint at lower fragmentation velocities, we performed additional simulations where the fragmentation velocity of dust grains is fixed at 1~${\rm m\,s}^{-1}$ throughout the whole disc. The outcome is shown in Figure \ref{fig:figA0}.

We find that planetesimals can form at the water ice line only in the case of a quiescent disc $(\alpha_t = 10^{-4})$ due to the larger Stokes number of the pebbles. Since the Stokes number ${\rm St} \propto u_{\rm frag}^2/\alpha_{\rm t}$ in the fragmentation limited regime \citep{Birnstiel2009}, dust grains of the same size will have a larger Stokes number in a quiescent disc than in a turbulent disc. This effect coupled with the low level of turbulence in the disc promotes the pile up of pebbles at the water ice line and leads to the formation of planetesimals. 

In contrast, small dust grains in a turbulent disc (e.g., $\alpha_t = 10^{-3}$) are transported away quickly by the gas, thus preventing pebbles from piling up at ice lines and consequently hindering the formation of planetesimals.

\section{Disc composition}
\label{sec:appendix_disc_composition}

\begin{table} [h!]
\centering
    \caption{Abundances of elements included in our model \citep{Asplund2009}.}
    \label{tab:elements}
    \begin{tabular}{c c}
    \hline\hline
    Element & Abundance\\ 
    \hline
    He/H & 0.085 \\
    O/H & $4.90\times10^{-4}$ \\
    C/H & $2.69\times10^{-4}$ \\
    Mg/H & $3.98\times10^{-5}$ \\
    Si/H & $3.24\times10^{-5}$ \\
    Fe/H & $3.16\times10^{-5}$ \\
    \hline
    \end{tabular}
\end{table} 

\begin{table}[h!]
\centering
    \caption{Molecules included in our model.}
    \label{tab:molecules}
    \begin{tabular}{c c c}
    \hline\hline
    Molecule & $T_{\rm cond}$~(K) & Volume mixing ratio \\ 
    \hline
    CO              & 20   & 0.2 $\times$ C/H \\
    CH$_4$          & 30   & 0.1 $\times$ C/H \\
    CO$_2$          & 70   & 0.1 $\times$ C/H \\
    H$_2$O          & 150  & O/H - (CO/H + 2 $\times$ CO$_2$/H +  \\
                    &      & 4 $\times$ Mg$_2$SiO$_4$/H + 3 $\times$ MgSiO$_3$/H) \\ 
    C               & 631  & 0.6 $\times$ C/H  \\
    Mg$_2$SiO$_4$   & 1354 & Mg/H - Si/H \\
    Fe              & 1357 & Fe/H \\
    MgSiO$_3$       & 1500 & Mg/H - 2 $\times$ (Mg/H - Si/H)  \\
    \hline
    \end{tabular}
\end{table}

\section{Planetesimal formation at ice lines in discs with super-solar metallicities}
\label{AppendixSupersolar}
\begin{figure*}
   \includegraphics[width=\textwidth]{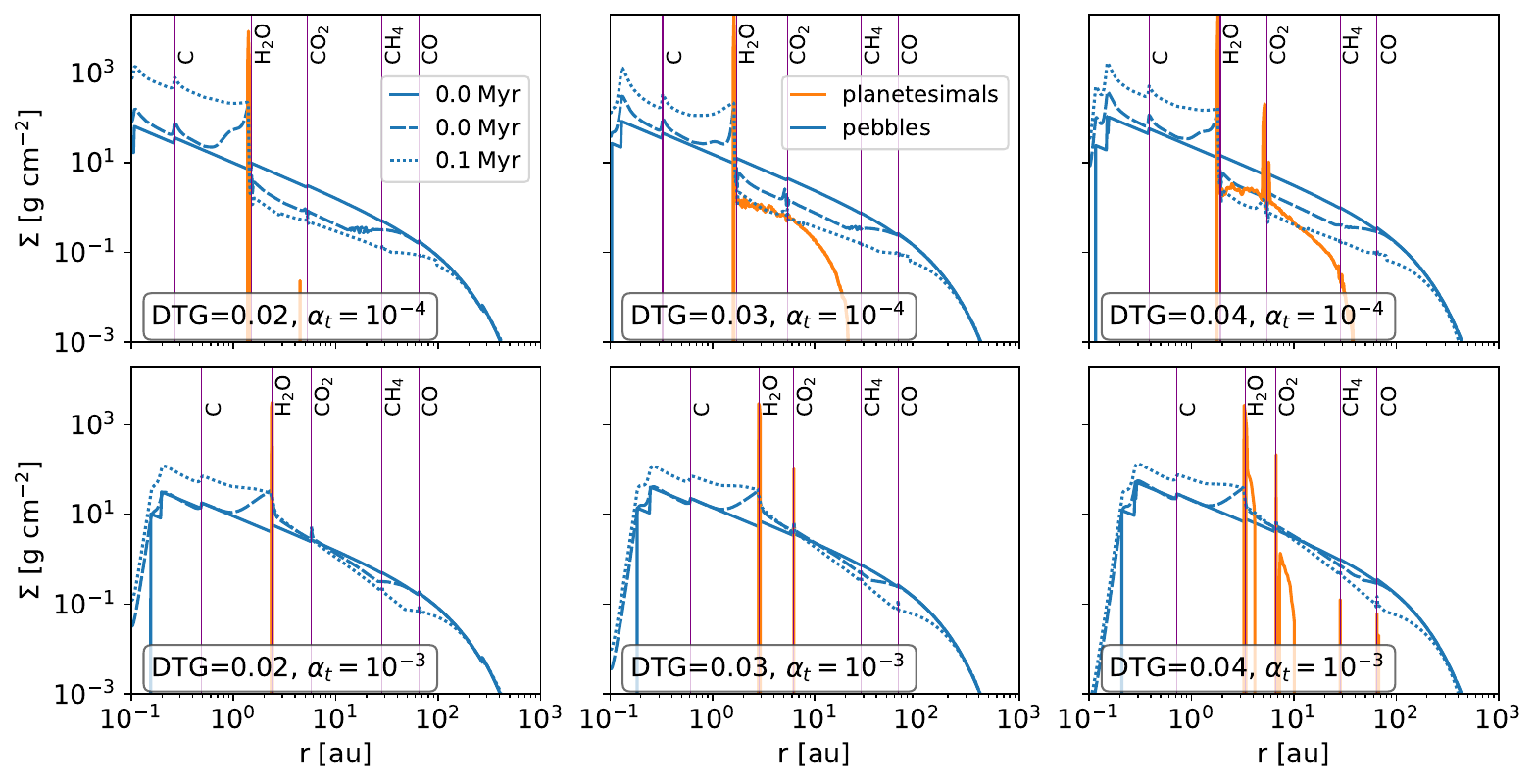}
   \caption{Planetesimal formation at ice lines in discs at super-solar metallicities. The top and bottom panels show simulations with $\alpha_{\rm t}=10^{-4}$ and $\alpha_{\rm t}=10^{-3}$, respectively, in a disc with $r_{\rm c}=100$ au.}
   \label{fig:figA}
\end{figure*}
We tested the possibility of planetesimal formation at super-solar metallicities at the ice lines of volatile species exterior to the water ice line. The results are shown in Figure \ref{fig:figA}, where we used dust-to-gas ratios 0.02, 0.03 and 0.04 corresponding to the scaled [Fe/H] values 0.2, 0.3 and 0.4, respectively.

The top panel of Figure \ref{fig:figA} shows the simulations in a quiescent disc with $\alpha_{\rm t}=10^{-4}$. Here, for the ice lines exterior to the water ice line, planetesimal formation only takes place at the CO$_2$ ice line, but this requires a dust-to-gas ratio of at least 0.02. For the case of dust-to-gas ratios of 0.03 and 0.04, planetesimal formation takes place in broader disc regions exterior to the water ice line because of the increased dust-to-gas ratios lead to increased traffic jam effect in the disc, in agreement with \cite{DrazkowskaAlibert2017}.

For the case of a turbulent disc with $\alpha_{\rm t}=10^{-3}$ in the bottom panel of Figure \ref{fig:figA}, planetesimal formation takes place at the CO$_2$ ice line requires a lower dust-to-gas ratios of 0.03 compared to the case of quiescent disc with $\alpha_{\rm t}=10^{-4}$. However, planetesimal formation in broader disc regions is still difficult for the case of $\alpha_{\rm t}=10^{-3}$ despite the large amounts of solid material in the disc. As explained before, because of high viscosity, dust grains drift more slowly which reduces the traffic jam effect, which hardly cascades beyond the ice lines compared to the quiescent disc. At the CH$_4$ and CO ice lines, traces of planetesimals form when the dust-to-gas is increased to 0.04. Consequently it seems that discs with higher viscosity only allow planetesimal formation at discrete locations, while discs with lower viscosities allow planetesimal formation in broader ranges of the disc. 

\section{How vertical settling affects planetesimal formation}
\label{AppendixVerticalSettling}
\begin{figure*}
   \includegraphics[width=\textwidth]{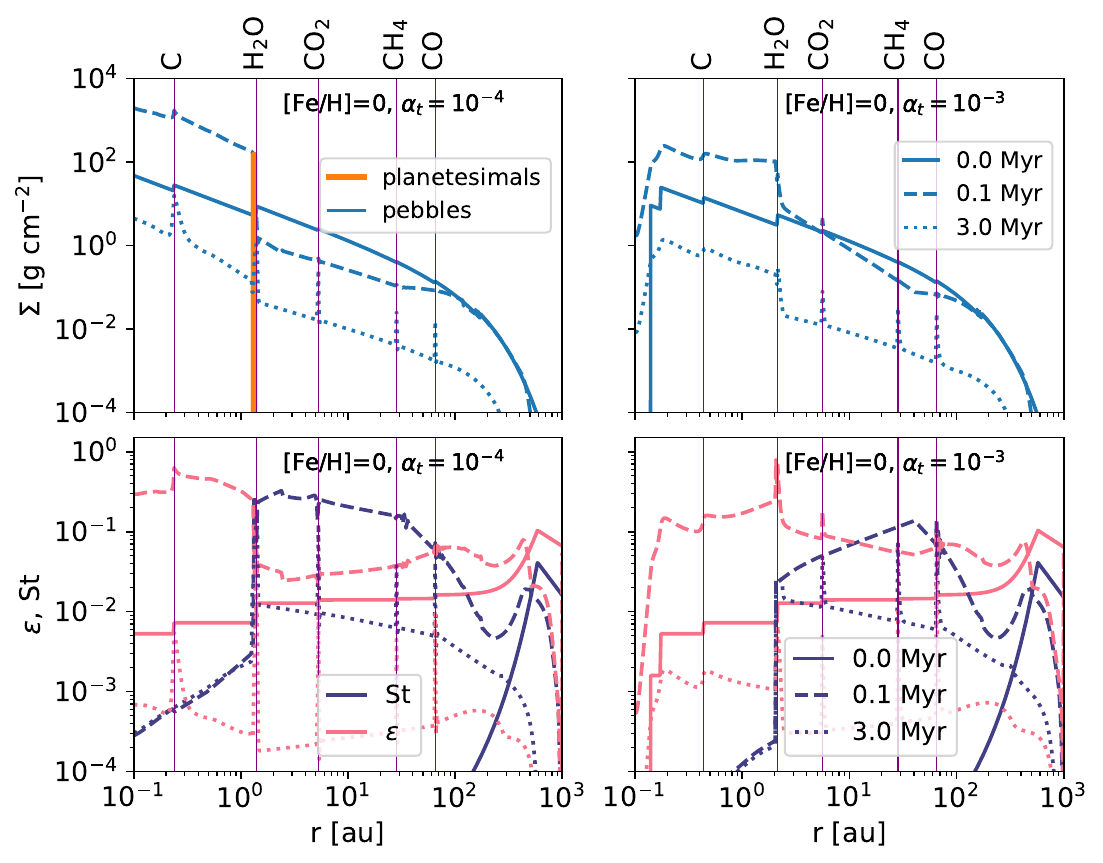}
   \caption{Planetesimal formation in a disc with vertical settling parameter $\alpha_{\rm z}=10^{-3}$, solar metallicity [Fe/H] $= 0$ and a disc size $r_{\rm c}=100$ au.}
   \label{fig:figB1}
\end{figure*}

\begin{figure*}
   \includegraphics[width=\textwidth]{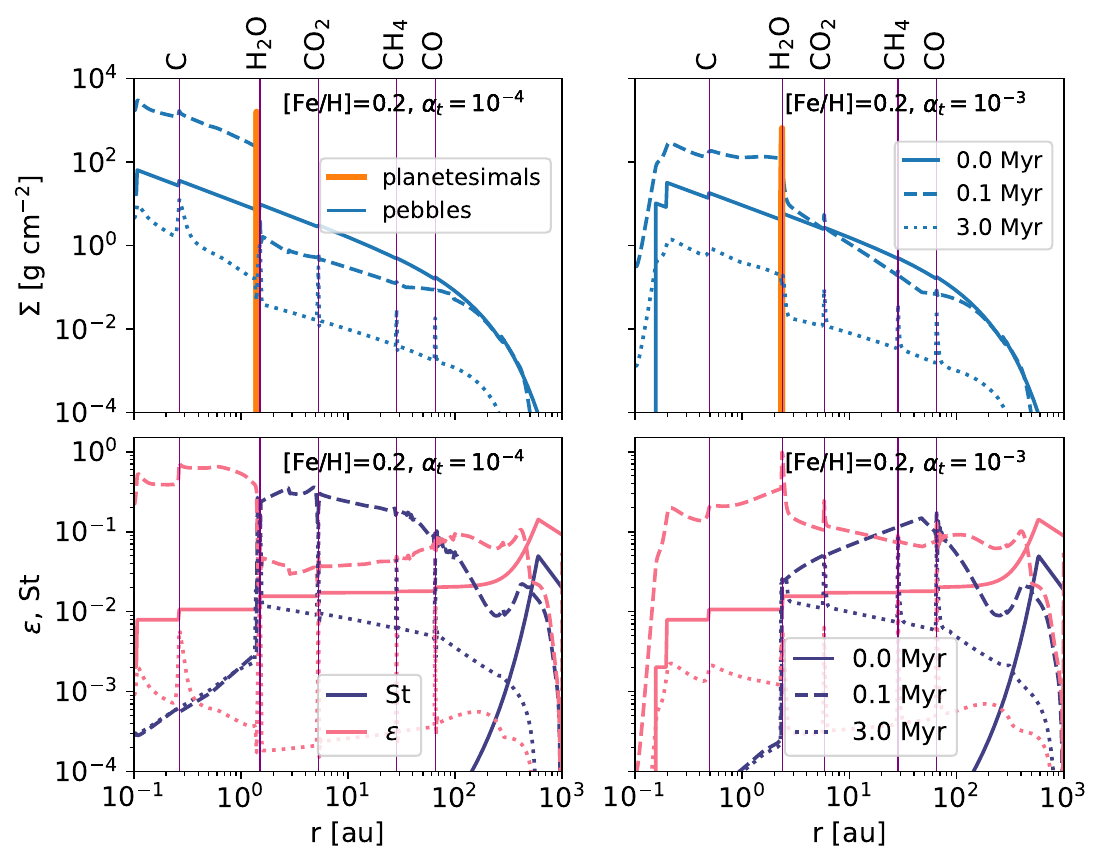}
   \caption{Planetesimal formation in a disc with vertical settling parameter $\alpha_{\rm z}=10^{-3}$, super-solar metallicity [Fe/H] $= 0.2$ and a disc size $r_{\rm c}=100$ au.}
   \label{fig:figB2}
\end{figure*}

The midplane dust-to-gas is influenced by the extent to which dust grains settle to the midplane through the vertical settling parameter $\alpha_{\rm z}$, which in turn affects the planetesimal formation efficiency.  Using vertical settling parameter $\alpha_{\rm z}=10^{-3}$, we tested how weak grain settling affects planetesimal formation at solar and super-solar metallicities as shown in Figures \ref{fig:figB1} and \ref{fig:figB2}, respectively.

From Figure \ref{fig:figB1}, planetesimal formation can still take place in a less turbulent disc with $\alpha_{\rm t}=10^{-4}$ compared to the turbulent disc with $\alpha_{\rm t}=10^{-3}$. This is because, in the less turbulent disc, the large size grains settle closer to the midplane compared to the more turbulent disc that produces smaller size grains which hang further above the midplane. This reduces the midplane dust-to-gas ratios, which makes planetesimal formation more difficult. However, increasing the disc metallicity to a super-solar value of [Fe/H] $= 0.2$ makes planetesimal formation in the more turbulent disc with $\alpha_{\rm t}=10^{-3}$ as shown in Figure \ref{fig:figB2} when the vertical settling parameter is set to $\alpha_{\rm z}=10^{-3}$. At super solar metallicity, there is enough solid material that increases the midplane dust-to-gas ratios which then supports planetesimal formation.

\end{appendix}

\end{document}